# Assessing the risks of "infodemics" in response to COVID-19 epidemics

Riccardo Gallotti[1], Francesco Valle[1], Nicola Castaldo[1], Pierluigi Sacco[2,3,4*] & Manlio De Domenico[1*]

[1]*CoMuNe Lab, Fondazione Bruno Kessler, Via Sommarive 18, Povo, 38123, Trento, Italy*

[2]*IULM University, Via Carlo Bo, 1, 20143 Milan, Italy*

[3]*Berkman-Klein Center for Internet & Society, Harvard University, 23 Everett St 2, Cambridge MA 02138 USA*

[4]*Fondazione Bruno Kessler, Via Santa Croce, 77, 38122 Trento, Italy*

\* Corresponding author: mdedomenico@fbk.eu, pierluigi_sacco@fas.harvard.edu

**Our society is built on a complex web of interdependencies whose effects become manifest during extraordinary events such as the COVID-19 pandemic, with shocks in one system propagating to the others to an exceptional extent. We analyzed more than 100 millions Twitter messages posted worldwide in 64 languages during the epidemic emergency due to SARS-CoV-2 and classified the reliability of news diffused. We found that waves of unreliable and low-quality information anticipate the epidemic ones, exposing entire countries to irrational social behavior and serious threats for public health. When the epidemics hit the same area, reliable information is quickly inoculated, like antibodies, and the system shifts focus towards certified informational sources. Contrary to mainstream beliefs, we show that human response to falsehood exhibits early-warning signals that might be mitigated with adequate communication strategies.**

Human societies build on social, economic, environmental and technological systems whose dynamics are inherently complex and often highly unpredictable in the short term. The effects of this deep-layered structural interdependency [1,2] become manifest during extraordinary events, such as natural catastrophes or pandemics, where shocks propagate across systems. Although the level of complexity of past human societies has been often underestimated, it can be claimed that, in the past few decades, the acceleration of globalization processes has brought about an unprecedented level of large-scale interdependencies, from trade of goods to communications, that dramatically changed the temporal scales of shock propagation. However, how to map and understand the potential diffusion pathways which might lead to major systemic crises or even collapse is still unknown.

The high levels of specialization [3] and adaptive flexibility [4] of human societies rely upon complex, multifaceted forms of cooperation, to the point of characterizing humans as a super-cooperator species [5]. One would therefore expect that the human propensity to cooperate would be further magnified when facing major threats that put collective wellbeing at risk. In large, complex societies, an important mediator of large-scale cooperation is communication [6], which may be crucial to coordinate individual perceptions and behaviors in the pursuit of the common interest [7]. The recent explosion of publicly shared, decentralized information production that characterizes digital societies [8] and in particular social media activity [9] provides an exceptional laboratory for the observation and the study of these complex social dynamics [10], and potentially functions as a very powerful resource to enact effective, pro-social cooperation and coordination in large-scale crises [11]. Global pandemics are certainly an instance of such crises, and the current outbreak of COVID-19 may therefore be thought of as a natural experiment to observe social responses to a major threat that may potentially escalate to catastrophic levels, and has already managed to seriously affect levels of economic activity, and radically alter human social behaviors across the globe.

In this study, we show that information dynamics tailored to alter individuals' perceptions and, consequently, their behavioral response, is able to drive collective attention [12] towards false [13,14] or inflammatory [15] content, a phenomenon named infodemics [16–19], sharing similarities with more traditional epidemics and spreading phenomena [20–22]. Contrary to what it could be expected in principle, what this natural experiment reveals is that, on the verge of a threatening global pandemic emergency due to SARS-CoV-2 [23–25], human communication activity is to a significant extent characterized by the intentional production of informational noise and even of misleading or false information [26]. This generates waves of unreliable and low-quality information with potentially very dangerous impacts on the social capacity to respond adaptively at all scales by rapidly adopting those norms and behaviors that may effectively contain the propagation of the epidemics. Spreading false information or even conspiracy theories that support implausible explanations of the causal forces at work behind the crisis may create serious confusion and even discourage people from taking the crisis seriously or responsibly, all the more so, the more such signals receive social validation and spread across social groups and communities [27]. Therefore, if on the one hand we face the risks of a global epidemics threat, requiring outstanding efforts for modeling and anticipating the time course of the spreading [25], on the other hand we can speak of an infodemics threat [28], where low-quality content provides an alternative for news consumption to unclear official communications. The infodemics can be thought, similarly to epidemics, as an outbreak of false rumors and fake news with unexpected effects on social dynamics (see Fig. 1). In fact, the dangerousness of infodemics can compare and sum up to a large extent to that of the epidemics itself [29].

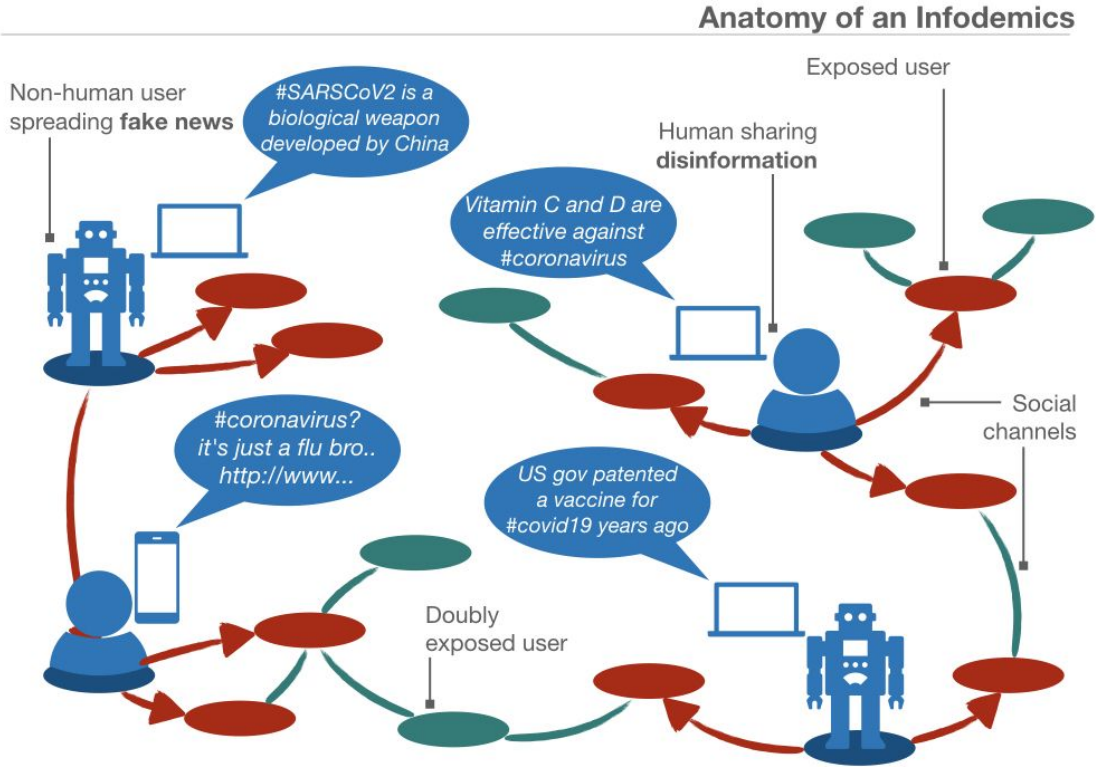

**Fig. 1: How infodemics works**. Human and non-human accounts forge unreliable content – such as fake or untrustworthy news – about the COVID19, a topic attracting the collective attention of the whole world. Their followers are exposed to such content, and reshare it, becoming infectious agents: infodemics realizes when multiple spreading processes co-occur. Some users might be exposed multiple times to the same content or to different contents generated by distinct accounts, as in epidemics spread.

As shown in Fig. 1, an infodemics is the result of the simultaneous action of multiple human and non-human sources of fake or unreliable news. As users are repeatedly hit by a given message from different sources, this works as an indirect validation of its reliability and relevance, leading the user to spread it in turn, and to become an informationally infectious agent.

The COVID-19 crisis allows us to provide a rigorous, evidence-based assessment of such risks, and of the real-time interaction of the infodemic and epidemic layers [21].

We focus our attention on the analysis of messages posted on a popular microblogging platform [30], an online social network characterized by heterogeneous connectivity [31] and topological shortcuts typical of small-world systems [32]. Information spreading on this type of networks is well understood in terms of global cascades in a population of individuals who have to choose between complementary alternatives, while accounting for the behavior and the relative size of their social neighborhood [33], and accounting for factors which characterize the popularity of specific content, like the memory time of users and the underlying connectivity structure [34]. However, the exact fundamental mechanisms responsible for the spread of false information and inflammatory content, e.g. during political events [15,35 36,37], remains fundamentally unknown. Recently, it has been suggested that this challenging phenomenon might be due to the fact that, at population level, the dynamics of multiple interacting contagions are indistinguishable from social reinforcement [38].

This peculiar feature suggests that infodemics of news consumption should be analyzed through the lens of epidemiology to gain insights about the role of human and non-human activities in spreading reliable as well as unreliable news. To this aim, we monitored the social media and collected more than 112 millions messages in 64 languages from around the world about COVID-19, between 21 January and 10 March 2020 (see Methods for details).

By using state-of-the-art machine learning techniques to analyze the online behavior of users (see Methods for details), we have discovered an extraordinary activity of automated agents, referred to as social bots [14,15,35,39]. Specifically, we estimate that 40.4% of online messages during this period were due to such automated agents, doubling the activity with respect to estimates of only four years ago [35].

Where available, we have extracted URLs from messages, collecting about 20.7 millions links (3.3 millions unique) pointing to websites external to the platform. Each URL is therefore used for fact-checking, inheriting the reliability of its source (see Methods). About 50% of URLs have been fact-checked by screening almost 4,000 expert-curated web domains, whereas the remaining corpus was pointing to disappeared web pages or to content not classifiable automatically (eg, videos on YouTube) and unpopular sources. This method allowed us to overcome the limitations due to text mining of different languages for the analysis of narratives.

To better understand the diffusion of these contents across countries, we have filtered messages with geographic information. About 0.84% of collected posts were geo-tagged by the user, providing highly accurate information about their geographic location. However, by geocoding the information available in users' profiles, we were able to extend the corpus of geolocated messages to about 56% of the total observed volume (see Methods). A total of more than 60 millions geolocated messages, containing more than 9 millions news have been analyzed. For each message, we have used an accurate machine learning approach to classify the author as human or non-human (i.e., bot), while keeping the distinction between verified and unverified users. Usually, verification is performed by the social platform to clearly identify accounts of public interest and certify they are authentic. The number of followers $K_u$ of a single user $u$ defines the exposure, in terms of potential visualizations at first-order approximation, of a single message $m$ posted by user $u$ at time $t$. Let $M_u(t, t + \Delta t)$ indicate the set of messages posted by user $u$ in a time window of length $\Delta t$. Since there are four different classes of users – namely verified bots (VB), unverified bots (UB), verified humans (VH) and unverified humans (UH) – we define the exposure due to a single class $C_i$ ($i = $ VB,UB,VH,UH) as

$$E_i(t, t+\Delta t) = \sum_{u \in C_i} \sum_{m \in M_u(t,t+\Delta t)} K_u \quad (1)$$

Note that different users of the same class might have overlapping social neighborhoods: those neighbors might be reached multiple times by the messages coming from distinct users of the same class, therefore our measure of exposure accounts for this effect. Note that our measure provides a lower bound to the number of exposed users, because we do not track higher-order transmission pathways: a user might adopt a content by reading it, while not resharing it. In this case there is no way to account for such users.

Finally, for each message, we identify the presence of links pointing to external websites: for each link we verify if it comes from a trustworthy source or not (see Methods). The reliability $r_m$ of a single message $m$ is either 0 or 1, because we discard all web links that can not be easily assessed, such as the ones shortened by third-party services that expired or point to unreachable destinations, and the ones pointing to external platforms, such as YouTube, where it is not possible to automatically classify the reliability of the content. The news reliability of messages produced by a specific class of users is therefore defined as

$$R_i(t, t+\Delta t) = \sum_{u \in C_i} \sum_{m \in M_u(t,t+\Delta t)} r_m \quad (2)$$

Unreliability can be defined similarly, replacing $r_m$ with $1-r_m$. Exposure and reliability are useful descriptors that, however, do not capture alone the risk of infodemics. For this reason we have developed an Infodemic Risk Index (IRI) which quantifies the rate at which a generic user is exposed to unreliable news produced by a specific class of users (partial IRI) or by any class of users (IRI):

$$pIRI_i(t, t+\Delta t) = \frac{\sum\limits_{u \in C_i} \sum\limits_{m \in M_u(t,t+\Delta t)} K_u(1-r_m)}{\sum\limits_{i} E_i(t, t+\Delta t)} \quad (3)$$

$$IRI(t, t+\Delta t) = \sum_{i} pIRI_i(t, t+\Delta t) \quad (4)$$

Both indices are well defined and range from 0 (no infodemic risk) to 1 (maximum infodemic risk). Note that we can calculate all the infodemics descriptors introduced above at a desired level of spatial and temporal resolution. IRI is robust to user classification, making it an indicator not sensitive to performances of bot detection algorithms.

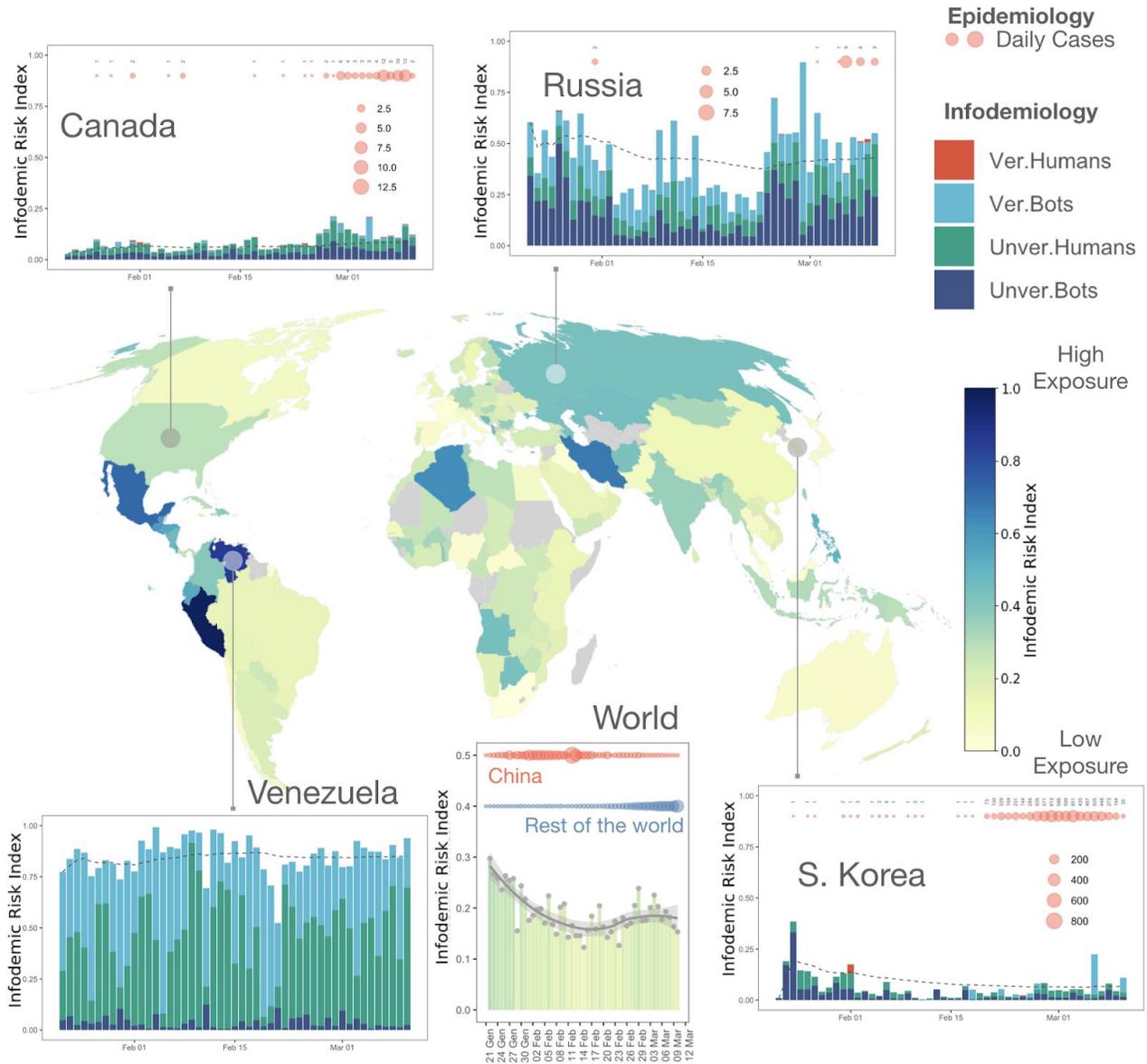

**Fig. 2: Mapping the infodemic risk worldwide.** The infodemic risk of each country, aggregated over time, is color-coded on the map. Panels show the evolution of the risk over time for some countries: bars indicate the contribution of each category (human/bot, verified/unverified) to the overall risk. The risk evolution for the whole world is also shown, demonstrating an overall decrease over time (bottom-middle panel). Markers horizontally aligned at the top of each panel indicate the daily confirmed epidemiological cases, with their number encoded by markers' size.

Figure 2 shows how countries characterized by different levels of infodemic risk present very different profiles of news sources. In a low-risk country such as South Korea, the level of infodemic risk remains small throughout apart from an isolated spike in the early phase. As the contagion spreads to significant levels, the infodemic risk further decreases, signalling an increasing focus of the public opinion toward reliable news sources. Canada presents a slightly higher level of infodemic risk, and unlike South Korea, we see that the risk level increases as the epidemics spread, but stays at low levels. At the opposite, in a high-risk country such as Venezuela, the infodemics is in full swing throughout the period of observation, and in addition to the expected activity from unverified sources one notices that even verified ones contribute to a large extent to the infodemics. The relationship with biological contagion patterns cannot be checked here due to lack of reliable data. Finally, in a relatively high-risk country such as Russia we notice that infodemic risk is erratic with sudden, very pronounced spikes, and again also verified sources play a major role. Here too, information about the epidemics is fragmented and mostly unreliable. Overall, the global level of infodemic risk tends to decrease as the epidemics spread globally, suggesting that evidence of the expansion of the contagion leads people to look for relatively more reliable sources, and that verified influencers with many followers started inoculating the system with more reliable news (see Supplementary Figures 3 and 4), playing a role that presents interesting analogies to that of antibodies in the treatment of an infectious disease. This overall pattern is confirmed in terms of measures of Infodemic Risk aggregated daily and at country level (Fig. 3 and Supplementary Fig. 5). The effect is particularly pronounced with the escalation of the epidemics, suggesting that this effect could be mediated by levels of perceived social alarm. It is also interesting to observe though that countries with high infodemic risk might also be more unreliable in terms of reporting of epidemic data, thus altering the perceptions of people and indirectly misleading them in their search for reliable information.

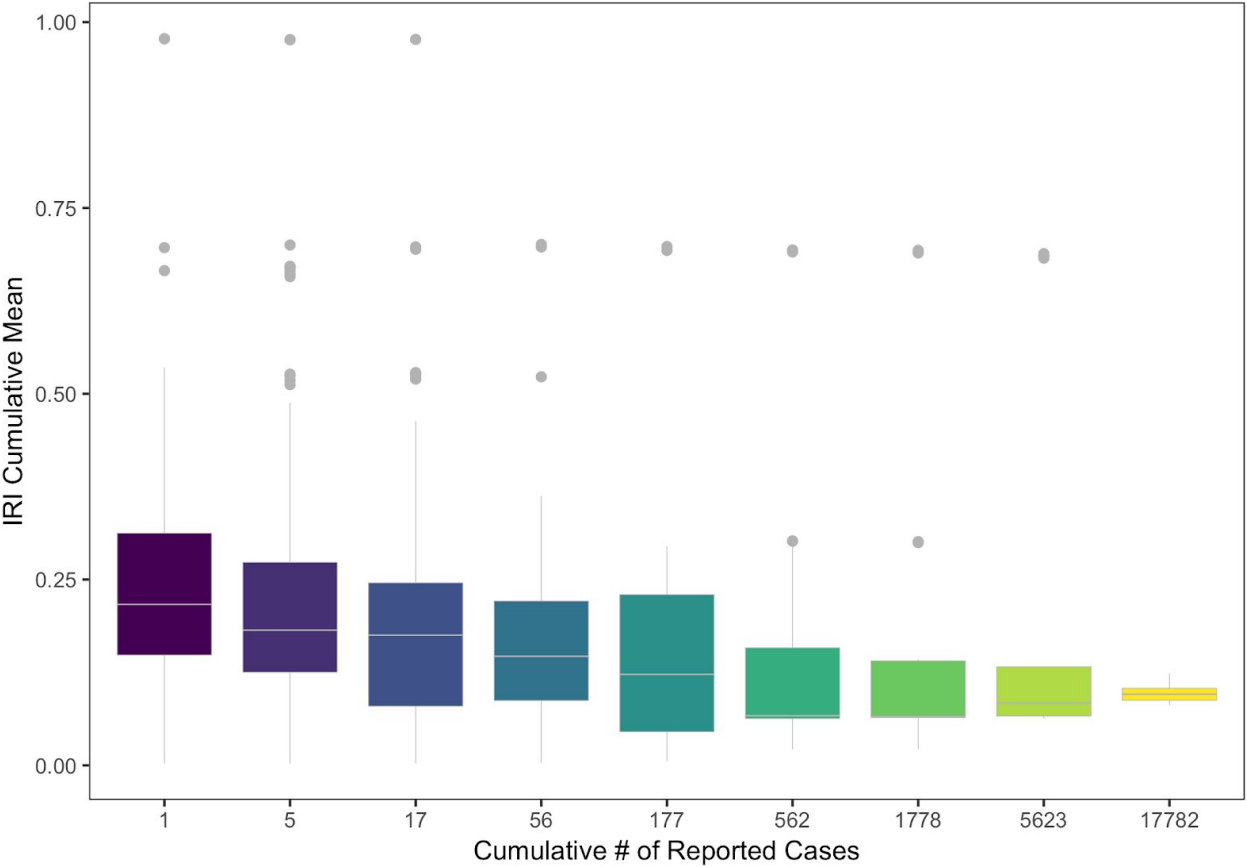

**Fig. 3: Positive behavioral response to infodemics after COVID-19 hits countries**. Aggregated view of the evolution of the infodemic risk index (IRI) for increasing number of reported cases. For each day and each of the 178 countries considered in our analysis, we compute the cumulative mean of the Infodemic Risk Index (computed as the cumulative sum of the IRI between the 22nd of January and specific day and the number of days in this date range). We aggregate days and countries with a similar cumulative number of reported cases, using bins of increasing size to compensate for the limited number of countries that reached high levels of contagion at the time of the analysis and reporting the average value in the x

axis. This allows us to describe, using boxplots, the drop in IRI as the number of cases grows in a country. Note that, in boxplots, the difference between two boxes is significant when corresponding middle lines lie outside of each other.

However, also the dynamic profiles of infodemic risk in countries with similar risk levels may be very different. Fig. 4 compares Italy with the United States. In the case of Italy the risk is mostly due to the activity of unverified sources, but we notice that with the outbreak of the epidemics, the production of misinformation literally collapses and there is a sudden shift to reliable sources. For the USA, misinformation is mainly driven by *verified* sources, and it remains basically constant even after the epidemics outbreak. Notice also how infodemic risk varies substantially across US states. As the USA lag significantly behind Italy in terms of the epidemics progression, it remains to be checked whether a similar readjustment is going to be observed for the USA later on. Fig. 4 shows, however, that the relationship between reduction of infodemic risk and expansion of the epidemics seems to be a rather general trend, as the relationship between number of confirmed cases and infodemic risk is (nonlinearly) negative, confirming the result shown in Fig. 3. Fig. 4 also shows how the evolution of infodemic risk among countries with both high message volume and significant epidemic contagion tends to be very asymmetric, with major roles played not only by countries such as Iran, but also United States, Germany, the Netherlands, Austria and Norway maintaining their relative levels, and other countries like Italy, South Korea and Japan significantly reducing it with the progression of the epidemics.

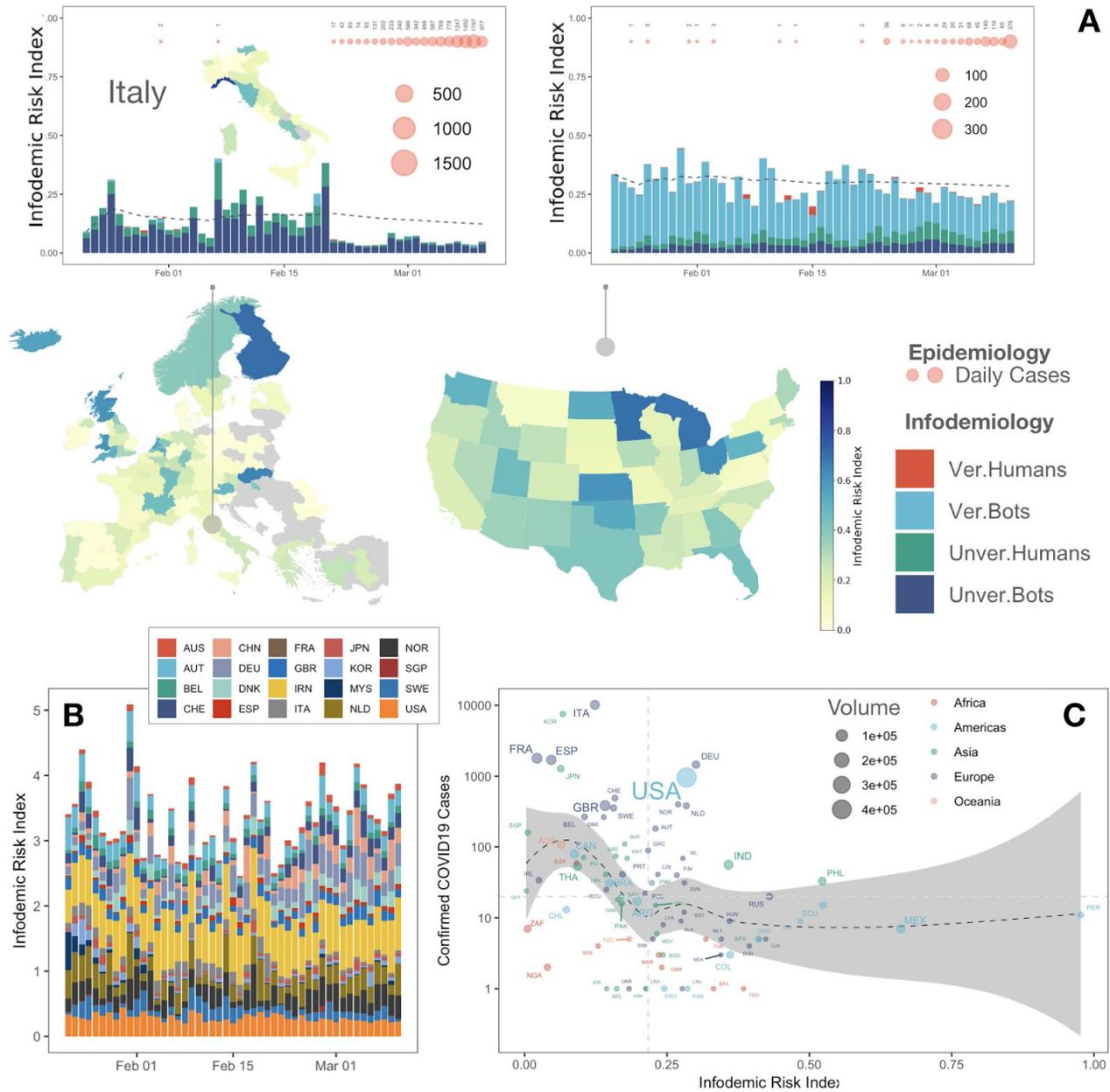

**Fig. 4: Infodemic evolution is country-dependent**. (A) as in Fig. 2, for European Union and USA at a finer resolution, with a detailed map for Italy (regional resolution). Areas with less than 10 messages, were excluded from the analysis and color-coded in grey. Note the striking drop in the Italian infodemic risk index coinciding with the first official report of non-imported epidemiological cases. (B) Risk evolution for countries characterized by a high volume of messages per day (at least one day with more than 2,000) and a high number of epidemiological cases (at least one day with more than 100). Countries respond in different ways to infodemics. (C) The number of epidemiological cases is shown against the infodemic risk index for all countries with at least one confirmed COVID19 case. Countries are colored by their Continent, with size proportional to the daily volume of messages generated. The shaded area and the dashed curve encode a local polynomial regression fit, here shown as a guide for the eye to highlight the highly nonlinear pattern relating epidemics and infodemics indices. China has to be considered as a major outlier due to its role in the global epidemic in terms of timing and size of the contagion, which makes it difficult to compare to other countries, and has therefore been removed from this analysis.

Our findings demonstrate that, in a highly digitalized society, the epidemic and the infodemic dimensions of a pandemic must be seen as two sides of the same coin. The infodemics is typically driven by the combined action of both human and non-human actors (bots), which pursue largely undisclosed goals. Perceived and actual biological and social risks feed upon one another, and may co-evolve in complex ways. Especially in situations where effective therapies to contrast the diffusion of the pandemic are not readily available, coordination of behaviors and diffusion of pro-social orientations driven by reliable information at all scales are the key resources for the mitigation of adverse effects. In this perspective, we can therefore think of an integrated public health approach where the biological and informational dimensions are equally recognized, taken into account, and managed through careful policy design. This could potentially include the birth of new, highly specialized professional figures such as that of the "infodemiologist".

Here, we have shown that in the context of the COVID-19 crisis, complex infodemic effects are indeed at work, with significant variations across countries, where level of socio-economic development is not the key discriminant to separate countries with high vs. low infodemic risk. In fact, we find that there are G9 countries with remarkable infodemic risk and developing countries with far lower risk levels. This means that, especially in countries where infodemic risk is high, the eventual speed and effectiveness of the containment of the COVID-19 could depend on a prompt regime switch in communication strategies and in the effective countervailing of the most active sources of the most dangerous categories of fake news. The escalation of the epidemics leads people to progressively pay attention to more reliable sources thus potentially limiting the impact of the infodemics, but the actual speed of adjustment may make a major difference in determining the social outcome, and in particular between a controlled epidemics and a global pandemics. This casts new light on the social mechanics of the infodemics-epidemics interaction, and may be of help to policy makers to design a more integrated strategic approach, by suitably embedding communication and information management into a comprehensive, extended public health perspective.

# References

1. Buldyrev, S. V., Parshani, R., Paul, G., Stanley, H. E. & Havlin, S. Catastrophic cascade of failures in interdependent networks. *Nature* **464,** 1025–1028 (2010).
2. Gao, J., Buldyrev, S. V., Stanley, H. E. & Havlin, S. Networks formed from interdependent networks. *Nat. Phys.* **8,** 40–48 (2011).
3. Cosmides, L., Barrett, H. C. & Tooby, J. Colloquium paper: adaptive specializations, social exchange, and the evolution of human intelligence. *Proc Natl Acad Sci USA* **107 Suppl 2,** 9007–9014 (2010).
4. Cortina, M. Adaptive flexibility, cooperation, and prosocial motivations: the emotional foundations of becoming human. *Psychoanalytic Inquiry* **37,** 436–454 (2017).

5. Nowak, M. & Highfield, R. *SuperCooperators: Altruism, Evolution, and Why We Need Each Other to Succeed*. (Simon and Schuster, 2011).

6. Miller, J. H., Butts, C. T. & Rode, D. Communication and cooperation. *J. Econ. Behav. Organ.* **47,** 179–195 (2002).

7. Balliet, D. Communication and Cooperation in Social Dilemmas: A Meta-Analytic Review. *Journal of Conflict Resolution* **54,** 39–57 (2010).

8. Benkler, Y. *The Wealth of Networks: How Social Production Transforms Markets and Freedom*. 528 (Yale University Press, 2006).

9. Fuchs, C. *Social media: A critical introduction*. (SAGE Publications Ltd, 2014). doi:10.4135/9781446270066

10. Giglietto, F., Rossi, L. & Bennato, D. The Open Laboratory: Limits and Possibilities of Using Facebook, Twitter, and YouTube as a Research Data Source. *J. Technol. Hum. Serv.* **30,** 145–159 (2012).

11. Ojo, A. & Mellouli, S. Deploying governance networks for societal challenges. *Gov. Inf. Q.* (2016). doi:10.1016/j.giq.2016.04.001

12. De Domenico, M. & Altmann, E. G. Unraveling the origin of social bursts in collective attention. *Sci. Rep.* **10,** 4629 (2020).

13. Vosoughi, S., Roy, D. & Aral, S. The spread of true and false news online. *Science* **359,** 1146–1151 (2018).

14. Shao, C. *et al.* The spread of low-credibility content by social bots. *Nat. Commun.* **9,** 4787 (2018).

15. Stella, M., Ferrara, E. & De Domenico, M. Bots increase exposure to negative and

inflammatory content in online social systems. *Proc Natl Acad Sci USA* **115,** 12435–12440 (2018).

16. Eysenbach, G. Infodemiology: The epidemiology of (mis)information. *Am. J. Med.* **113,** 763–765 (2002).

17. Eysenbach, G. Infodemiology and infoveillance: framework for an emerging set of public health informatics methods to analyze search, communication and publication behavior on the Internet. *J. Med. Internet Res.* **11,** e11 (2009).

18. Eysenbach, G. Infodemiology and infoveillance tracking online health information and cyberbehavior for public health. *Am. J. Prev. Med.* **40,** S154-8 (2011).

19. Zarocostas, J. How to fight an infodemic. *Lancet* **395,** 676 (2020).

20. Pastor-Satorras, R., Castellano, C., Van Mieghem, P. & Vespignani, A. Epidemic processes in complex networks. *Rev. Mod. Phys.* **87,** 925–979 (2015).

21. De Domenico, M., Granell, C., Porter, M. A. & Arenas, A. The physics of spreading processes in multilayer networks. *Nat. Phys.* **12,** 901–906 (2016).

22. Brockmann, D. & Helbing, D. The hidden geometry of complex, network-driven contagion phenomena. *Science* **342,** 1337–1342 (2013).

23. Huang, C. *et al.* Clinical features of patients infected with 2019 novel coronavirus in Wuhan, China. *Lancet* **395,** 497–506 (2020).

24. Zhu, N. *et al.* A Novel Coronavirus from Patients with Pneumonia in China, 2019. *N. Engl. J. Med.* **382,** 727–733 (2020).

25. Chinazzi, M. *et al.* The effect of travel restrictions on the spread of the 2019 novel coronavirus (COVID-19) outbreak. *Science* (2020). doi:10.1126/science.aba9757

26. Lazer, D. M. J. *et al.* The science of fake news. *Science* **359,** 1094–1096 (2018).

27. Waszak, P. M., Kasprzycka-Waszak, W. & Kubanek, A. The spread of medical fake news in social media – The pilot quantitative study. *Health Policy and Technology* **7,** 115–118 (2018).

28. Leung, G. M. & Leung, K. Crowdsourcing data to mitigate epidemics. *The Lancet Digital Health* (2020). doi:10.1016/S2589-7500(20)30055-8

29. Brainard, J., Hunter, P. R. & Hall, I. R. An agent-based model about the effects of fake news on a norovirus outbreak. *Rev Epidemiol Sante Publique* (2020). doi:10.1016/j.respe.2019.12.001

30. Kwak, H., Lee, C., Park, H. & Moon, S. What is Twitter, a social network or a news media? in *Proceedings of the 19th international conference on World wide web - WWW '10* 591 (ACM Press, 2010). doi:10.1145/1772690.1772751

31. Barabasi, A. L. & Albert, R. Emergence of scaling in random networks. *Science* **286,** 509–512 (1999).

32. Watts, D. J. & Strogatz, S. H. Collective dynamics of "small-world" networks. *Nature* **393,** 440–442 (1998).

33. Watts, D. J. A simple model of global cascades on random networks. *Proc Natl Acad Sci USA* **99,** 5766–5771 (2002).

34. Gleeson, J. P., O'Sullivan, K. P., Baños, R. A. & Moreno, Y. Effects of network structure, competition and memory time on social spreading phenomena. *Phys. Rev. X* **6,** 021019 (2016).

35. Bessi, A. & Ferrara, E. Social bots distort the 2016 U.S. Presidential election online

discussion. *FM* **21,** (2016).

36. Aral, S. & Eckles, D. Protecting elections from social media manipulation. *Science* **365,** 858–861 (2019).

37. Stella, M., Cristoforetti, M. & De Domenico, M. Influence of augmented humans in online interactions during voting events. *PLoS ONE* **14,** e0214210 (2019).

38. Hébert-Dufresne, L., Scarpino, S. V. & Young, J.-G. Macroscopic patterns of interacting contagions are indistinguishable from social reinforcement. *Nat. Phys.* (2020). doi:10.1038/s41567-020-0791-2

39. Ferrara, E., Varol, O., Davis, C., Menczer, F. & Flammini, A. The rise of social bots. *Commun. ACM* **59,** 96–104 (2016).

40. Pfeffer, J., Mayer, K. & Morstatter, F. Tampering with Twitter's Sample API. *EPJ Data Sci.* **7,** 50 (2018).

41. Ferrara, E. Disinformation and social bot operations in the run up to the 2017 French presidential election. *FM* **22,** (2017).

# Methods

### Data collection

We have followed a consolidated strategy for collecting social media data. We focused on Twitter, which is well-known for providing access to publicly available messages upon specific requests through their application programming interface (API). We have identified a set of hashtags and keywords gaining special collective attention, namely: *coronavirus, ncov, #Wuhan, covid19, covid-19, sarscov2, covid*. This set includes the official name of the virus and the disease, including the preliminary ones, as well as the name of the city of the first epidemic outbreak. We have used the Filter API – to collect the data in real time from 24 Jan 2020 to 10 Mar 2020 – and of the Search API – to collect the data between 21 Jan 2020 and 24 Jan 2020. Our choice allowed us to monitor, without interruptions and regardless of the language, all the tweets posted about COVID19 since when China reported more than 6,000 cases (20 Jan 2020), calling for the attention of the international community. The Stream API has the advantage of providing *all* the messages satisfying our selection criteria and posted to the platform in the period of observation, provided that their volume is not larger than 1% of the overall – unfiltered – volume of posted messages. Above 1% of the overall flow of information, the Filter API provides a sample of filtered tweets and communicates an estimate of the amount of lost messages. Note that this choice is the safest as to date: in fact, it has been recently shown that biases affecting Sample API (which samples data based on rate limits), for instance, are not found in REST and Filter APIs [40].

We estimate that until 24 Feb 2020 we lost about 60,000 tweets out of millions, capturing more than 99.5% of all messages posted (see Supplementary Fig. 1). The global attention towards COVID19 increased the volume of messages after 25 Feb 2020: however, Twitter restrictions allowed us to get no more than 4.5 millions messages per day, on average. We have estimated a total of 161.2 millions tweets posted until 10 Mar 2020: we have successfully collected 112.6 millions of them, providing an unprecedented opportunity for infodemics analysis.

**Human vs non-human classification**

The classification of users into humans and non-humans (ie, bots) is based on machine learning. It is based on a well established algorithm based on deep learning [37] with state-of-the-art accuracy [15,41]. More in detail, our method has the highest accuracy (>90%) and precision in identifying bots (>95%) when compared with state-of-the-art methods. Our deep neural network model has the advantage to be more stable in the classification of certain users playing the role of broadcasters. Note that in this study we are making an explicit difference between verified and unverified human/non-human users. In fact, verified users should be considered as more authentic than unverified ones, because Twitter makes use of strict criteria for verification. Therefore, verified bot accounts might be broadcasters (whose behavior is manifestly different from the average behavior of a single human) or, in some cases, even celebrities and any case where it is very likely that the account is managed automatically and exhibits a non-human classical behavior.

**Fact Checking**

We have collected manually-checked web domains from multiple publicly available databases, including scientific and journalistic ones. Specifically, we have considered data shared by:
- M. Zimdar for the Washington Post (2016). https://www.washingtonpost.com/posteverything/wp/2016/11/18/my-fake-news-list-went-viral-but-made-up-stories-are-only-part-of-the-problem/
- C. Silverman for BuzzFeed News (2017). https://www.buzzfeednews.com/article/craigsilverman/inside-the-partisan-fight-for-your-news-feed
- Fake News Watch (2015). https://web.archive.org/web/20180213181029/http://www.fakenewswatch.com/
- PolitiFact (2017). https://www.politifact.com/article/2017/apr/20/politifacts-guide-fake-news-websites-and-what-they/

- Bufale.net (2018). https://www.bufale.net/the-black-list-la-lista-nera-del-web/
- Starbird et al, ICWSM (2018)
- Fletcher et al, Factsheets, Reuters Institute and U. of Oxford (2018). https://reutersinstitute.politics.ox.ac.uk/our-research/measuring-reach-fake-news-and-online-disinformation-europe
- Grinberg et al, Science 363, 374 (2019)
- MediaBiasFactCheck (2020). https://mediabiasfactcheck.com/

However, databases adopted different labeling schemes to classify web domains, therefore we first had to develop a unifying classification scheme, reported in the table below, and map all existing categories to a unique set of categories. Note that we have also mapped those categories to a coarse-grain classification scheme, distinguishing just between *reliable* and *unreliable*.

| Category | Harm Score | Type | Description |
|---|---|---|---|
| **SCIENCE** | 1 | RELIABLE | Domains providing content validated by scientific scrutiny |
| **MAINSTREAM MEDIA** | 2 | RELIABLE | Domains providing content that generally goes through professional fact checking and generally abiding by the rules of media accountability |
| **SATIRE** | 3 | UNRELIABLE | Domains providing content and is intentionally and explicitly aiming at providing a distorted representation of events as a form of humor and/or social critique |
| **CLICKBAIT** | 4 | UNRELIABLE | Domain providing content that generally distorts or intentionally misrepresents information to capture attention |
| **OTHER** | 5 | UNKNOWN | Domains pointing to general content that can not be easily classified, such as videos on YouTube. |
| **SHADOW** | 6 | UNKNOWN | Domains related to URL shortening, that can be classified *a priori*. We follow these URLs to get the unshortened URLs and assign this category only when unshortening is not successful. |
| **POLITICAL** | 7 | UNRELIABLE | Domains providing content that present a |
| | | | partisan representation and interpretation of facts to support a political position at the expense of rival ones. |
| **FAKE/HOAX** | 8 | UNRELIABLE | Domains providing manipulative and fabricated content with the purpose of misleading the public opinion on socially relevant issues and to provoke inflammatory responses |
| **CONSPIRACY/JUNK SCI** | 9 | UNRELIABLE | Domains providing systematically manipulative and fabricated content with the purpose of legitimizing implausible conceptualizations of facts and knowledge through argumentative methods that coarsely mimic those of scientific reasoning, generally targeting individuals or social groups as covert perpetrators of conspiracies or harmful actions |

We have found a total of 4,988 domains, reduced to 4,417 after removing hard duplicates across databases. Note that a domain is considered a hard duplicate if its name and its classification coincides across databases.

A second level of filtering is applied to domains which are classified differently across databases (e.g., xyz.com might be classified as FAKE/HOAX in a database and as SATIRE in another database). To deal with these cases, we have adopted our own expert classification, by assigning to each category a *Harm Score* between 1 and 9. When two or more domains are soft duplicates, we keep the classification with the highest Harm Score, as a conservative choice. This phase of processing reduced the overall database to unique 3,920 domains.

The Harm Score classifies sources in terms of their potential contribution to the manipulative and mis-informative character of an infodemic. As a general principle, the more systematic and intentionally harmful the knowledge manipulation and data fabrication, the higher the Harm Score (HS). *Scientific content* has the lowest level of HS due to the rigorous process of validation carried out through scientific methods. *Mainstream media* content has the second lowest level of HS due to its constant scrutiny in terms of fact checking and media accountability. *Satire* is an unreliable source of news but due to its explicit goal of distorting or mis-representing information according to the specific cultural codes of humor and social critique, is generally identified with ease as an unreliable source. *Clickbait* is a more dangerous source (and thus ranking higher in HS) due to its intent to pass fabricated or mis-represented information and facts for true, with the main purpose of attracting attention and online traffic, that is, for mostly commercial purposes, but without a clear ideological intent. *Other* is a general purpose category that contains diverse forms of (possibly) misleading or fabricated content, not easily classifiable but likely including bits of ideologically structured content pursuing systematic goals of social manipulation, and thus ranking higher in HS. *Shadow* is a similar category to the previous one, where in addition links are anonymized and often temporary, thereby adding an extra element of unaccountability and manipulation that translates into a higher level of HS. *Political* is a category where we find an ample spectrum of content with varying levels of distortion and manipulation of information, also including mere selective reporting and omission, whose goal is that of building consensus for a political position against others, and therefore directly aiming at polluting the public discourse and opinion making, with a comparatively higher level of HS with respect to the previous categories. *Fake/hoax* contains entirely manipulated or fabricated inflammatory content which is intended to be perceived as realistic and reliable and whose goal may also be political, but fails to meet the basic rules of plausibility and accountability, thus reaching a even higher level of HS. Finally, the highest level of HS is associated to *conspiracy/junk science*, that is, to strongly ideological, inflammatory content that aims at building conceptual paradigms that are entirely alternative and oppositional to tested and accountable knowledge and information, with the intent of building self-referential bubbles where fidelized audiences are simply refusing a priori any kind of knowledge or information that is not legitimized by the alternative source itself or by recognized affiliates, as it is typical in sects of religious or other nature.

A third level of filtering concerned poorly defined domains, e.g., the ones explicitly missing top-level domain names, such as .com .org etc, as well as the domains not classifiable with our proposed scheme. This action reduced the database to the final number of 3,892 entries, whose statistics are reported in the tables below (see also Supplementary Fig. 2).

| Category | Count |
|---|---|
| **CLICKBAIT** | 47 |
| **CONSPIRACY/JUNK SCI** | 426 |
| **FAKE/HOAX** | 917 |
| **MSM** | 1284 |
| **OTHER** | 160 |
| **POLITICAL** | 697 |
| **SATIRE** | 177 |
| **SCIENCE** | 150 |
| **SHADOW** | 34 |
| Total | 3892 |

| Type | Count |
|---|---|
| **RELIABLE** | 1434 |
| **UNRELIABLE** | 2264 |
| **UNKNOWN** | 194 |
| Total | 3892 |

**Data availability**
The datasets generated during the current study are available from the corresponding author on reasonable request. Aggregated information, compliant with all privacy regulations on this matter, are publicly available online at the Infodemics Observatory (http://covid19obs.fbk.eu/) and on a permanent repository (Zenodo address/DOI will be with the publication of this manuscript).

**Acknowledgements**
We acknowledge the support of the FBK's Digital Society Department and the FBK's Flagship Project CHUB (Computational Human Behavior). We thank all FBK's Research Units for granting a privileged access to extraordinary high performance computing for the analysis of massive infodemics data. We thank Jason Baumgartner for kindly sharing data between 21 Jan and 24 Jan 2020.

**Authors' contributions**
M.D. conceived the study. M.D. and F.V. collected the data. R.G. and N.C. performed all experiments and analysed the data. M.D. and P.S. wrote the manuscript.

**Competing interests**
The authors declare no competing interests.

# Supplementary Figures

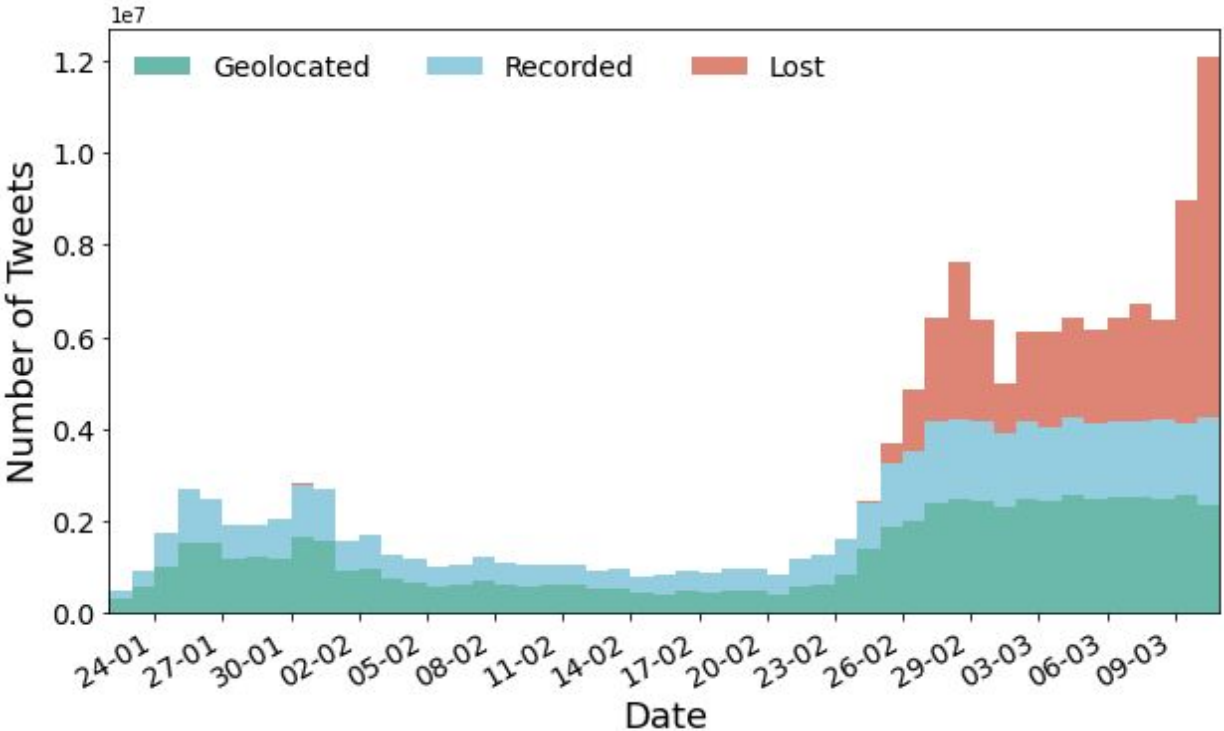

**Supplementary Fig. 1:** The evolution over time of the Twitter activity about the COVID19 pandemic (see Methods). We can observe a first increase in collective attention after the outbreak in Wuhan, China (between 24 Jan and 02 Feb 2020) and a second strong rise after the epidemics began to spread in northern Italy (20 Feb 2020 onwards). The fraction of Geoolocated (messages with shared locations, or geonamed, indicated in green) is constantly about 56% of the total volume recorded (indicated in blue). From 26 Feb, we reached the limit of the fraction of data shared by Twitter (see Methods), missing an increasing fraction of Tweets (indicated in red).

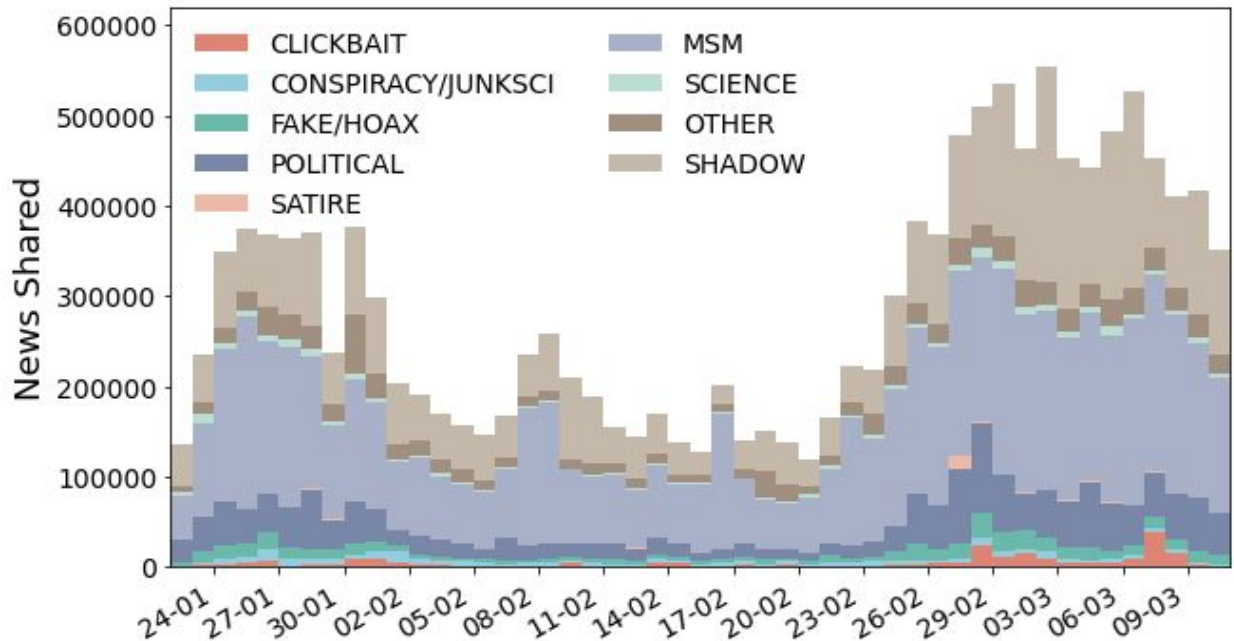

**Supplementary Fig. 2:** Temporal distribution of news shared on Twitter about COVID19, stratified by the category used in the fact-checking stage (see Methods). OTHER indicates URLs which point to general content (like YouTube videos), while SHADOW indicates shortened URLs which could not be unshortened (e.g., because pointing to removed web pages). Reliable news includes MSM and SCIENCE, whereas unreliable news includes the remaining categories. This analysis demonstrates that reliable sources are more represented than unreliable ones: however, they circulate in different ways and reach different targets, a feature that is perfectly captured by the infodemic risk index introduced in this study.

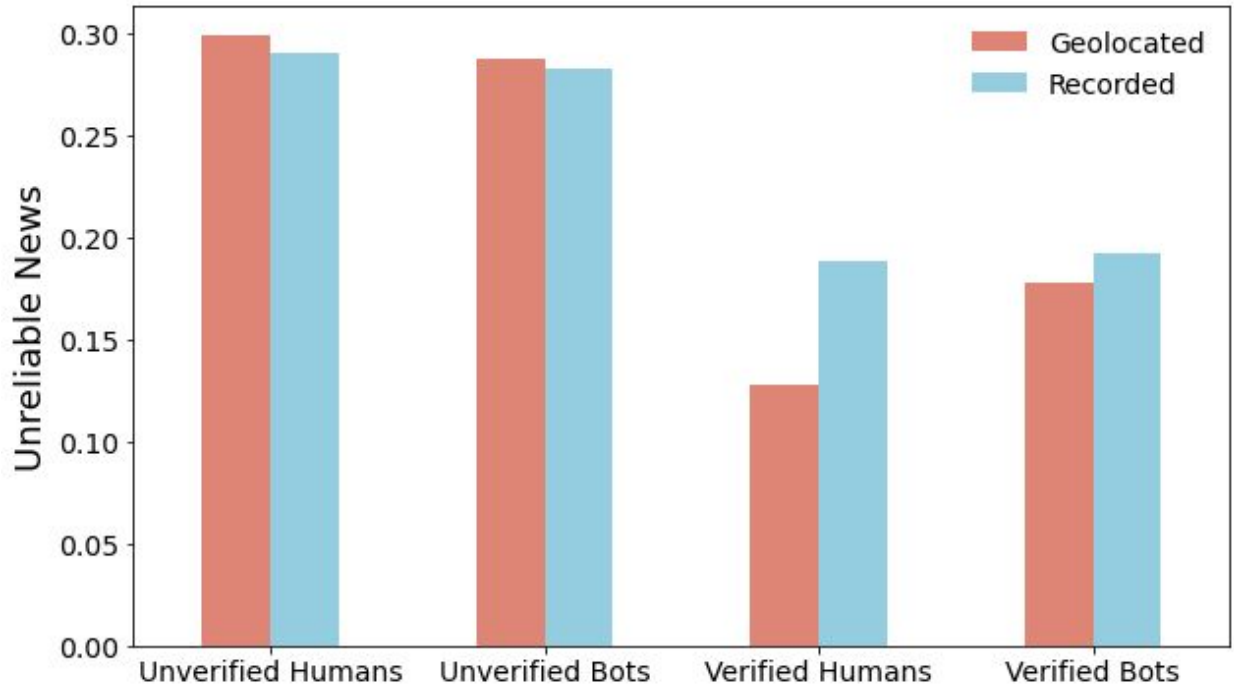

**Supplementary Fig. 3:** The fraction of unreliable news shared by the different classes of accounts. Unverified users, both humans and bots, share a larger fraction of unreliable news than verified users. We observe some differences between the values estimated on the Geolocated data used in this paper (red), and the values associated with the whole sample of Tweets recorded (blue). The difference, possibly caused by the correlation of this kind of behaviour with the choice of how customising the field "location" of one's profile, corresponds to a slight overestimate for unverified accounts and an underestimate of the fraction of fake news for verified profiles. As verified profiles participate in the risk index more strongly, we can expect the estimate of risk to be only greater if we could include all users.

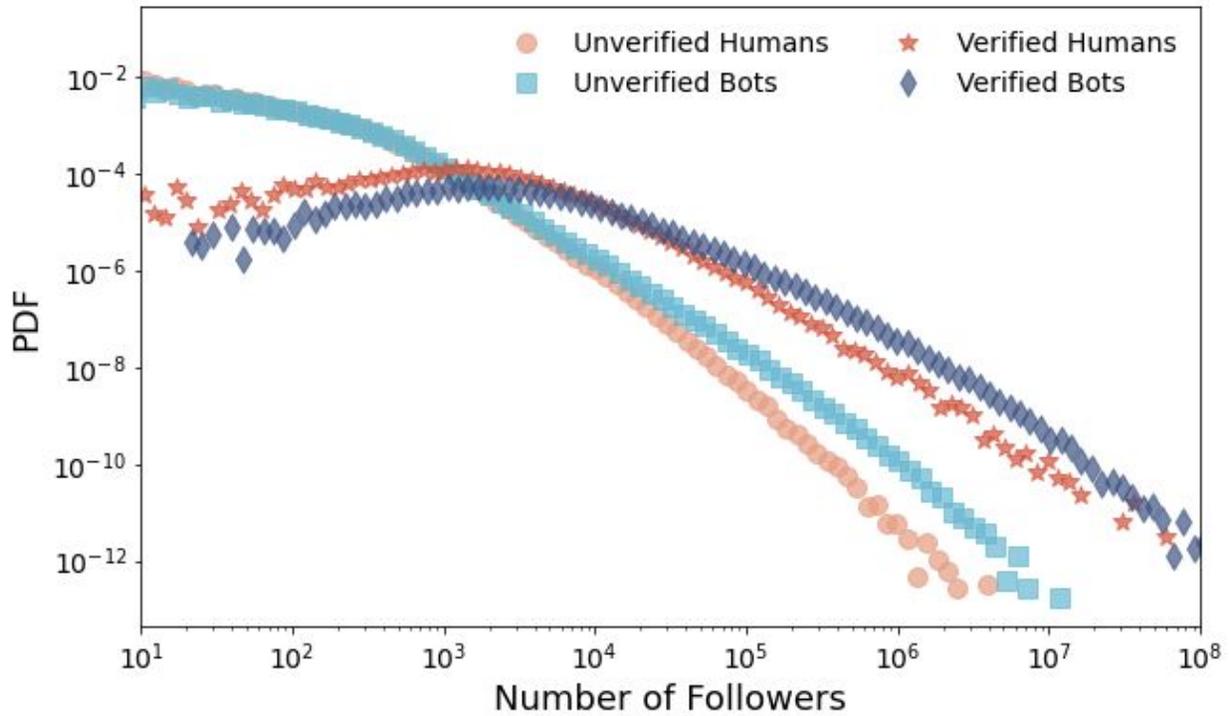

**Supplementary Fig. 4:** The probability distribution of the number of followers for the four classes of users considered in this study. All distributions display a fat-tail, but different categories of users have a different outreach. The unverified profiles have a significantly smaller number of followers than the verified ones. At the same time, profiles identified as bots have a larger number of followers. The average values are: 660 for Unverified Humans (circles), 1400 for Unverified Bots (squares, 51k for Verified Humans (stars) and 240k for Verified Bots (diamonds).

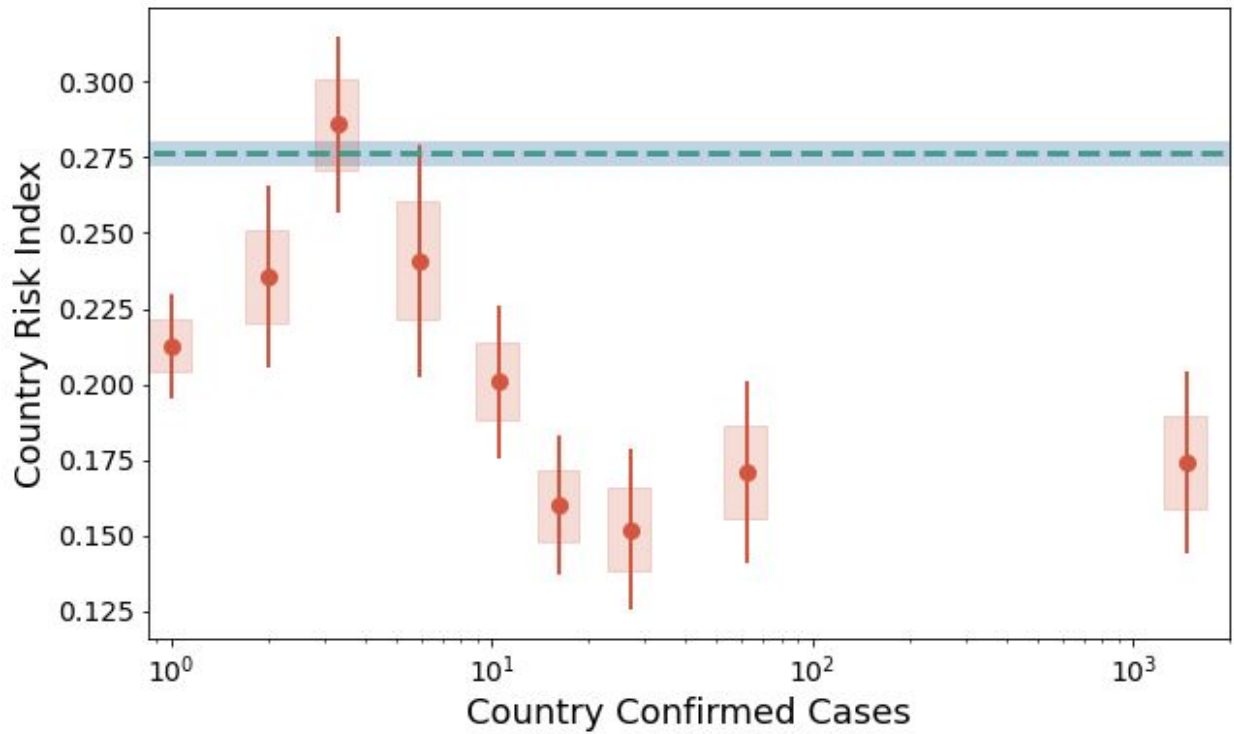

**Supplementary Fig. 5:** A second aggregated view of the evolution of risk index for increasing number of confirmed cases. Differently from Fig. 3, here we compute the average risk index for a single day in each country (instead of the cumulative value), for a total of 177 countries excluding China the origin of the COVID19 outbreak. We first compute the average value for days with no confirmed cases, which is 0.276 (blue line in figure). We then aggregate all days with confirmed cases in homogeneous bins and compute the average values (red points). We observe that reporting of the first case is associated with a drop in the risk index, followed by an increase back to the "natural level" in correspondence of 3 confirmed cases. Then, as the epidemics gain strength, the infodemic risk index decreases again, confirming what observed in Fig. 3. All shaded areas represent the s.e.m of the average risk index, error bars encode the 95% confidence intervals.